\documentstyle[preprint,tighten,prc,aps,epsf]{revtex}

\begin{document}
\draft

\def\simge{\hspace*{0.2em}\raisebox{0.5ex}{$>$}
     \hspace{-0.8em}\raisebox{-0.3em}{$\sim$}\hspace*{0.2em}}
\def\simle{\hspace*{0.2em}\raisebox{0.5ex}{$<$}
     \hspace{-0.8em}\raisebox{-0.3em}{$\sim$}\hspace*{0.2em}}
\def\bra#1{{\langle#1\vert}}
\def\ket#1{{\vert#1\rangle}}
\def\coeff#1#2{{\scriptstyle{#1\over #2}}}
\def\undertext#1{{$\underline{\hbox{#1}}$}}
\def\hcal#1{{\hbox{\cal #1}}}
\def\sst#1{{\scriptscriptstyle #1}}
\def\eexp#1{{\hbox{e}^{#1}}}
\def\rbra#1{{\langle #1 \vert\!\vert}}
\def\rket#1{{\vert\!\vert #1\rangle}}
\def\lsim{{ <\atop\sim}}
\def\gsim{{ >\atop\sim}}
\def\nubar{{\bar\nu}}
\def\psibar{{\bar\psi}}
\def\Gmu{{G_\mu}}
\def\alr{{A_\sst{LR}}}
\def\wpv{{W^\sst{PV}}}
\def\evec{{\vec e}}
\def\notq{{\not\! q}}
\def\notl{{\not\! \ell}}
\def\notk{{\not\! k}}
\def\notp{{\not\! p}}
\def\notpp{{\not\! p'}}
\def\notder{{\not\! \partial}}
\def\notcder{{\not\!\! D}}
\def\notA{{\not\!\! A}}
\def\notv{{\not\!\! v}}
\def\Jem{{J_\mu^{em}}}
\def\Jana{{J_{\mu 5}^{anapole}}}
\def\nue{{\nu_e}}
\def\mn{{M_\sst{N}}}
\def\mns{{M^2_\sst{N}}}
\def\me{{m_e}}
\def\mes{{m^2_e}}
\def\mq{{m_q}}
\def\mqs{{m_q^2}}
\def\mz{{M_\sst{Z}}}
\def\mzs{{M^2_\sst{Z}}}
\def\ubar{{\bar u}}
\def\dbar{{\bar d}}
\def\sbar{{\bar s}}
\def\qbar{{\bar q}}
\def\sstw{{\sin^2\theta_\sst{W}}}
\def\gv{{g_\sst{V}}}
\def\ga{{g_\sst{A}}}
\def\pv{{\vec p}}
\def\pvs{{{\vec p}^{\>2}}}
\def\ppv{{{\vec p}^{\>\prime}}}
\def\ppvs{{{\vec p}^{\>\prime\>2}}}
\def\qv{{\vec q}}
\def\qvs{{{\vec q}^{\>2}}}
\def\xv{{\vec x}}
\def\xpv{{{\vec x}^{\>\prime}}}
\def\yv{{\vec y}}
\def\tauv{{\vec\tau}}
\def\sigv{{\vec\sigma}}
\def\sst#1{{\scriptscriptstyle #1}}
\def\gpnn{{g_{\sst{NN}\pi}}}
\def\grnn{{g_{\sst{NN}\rho}}}
\def\gnnm{{g_\sst{NNM}}}
\def\hnnm{{h_\sst{NNM}}}

\def\xivz{{\xi_\sst{V}^{(0)}}}
\def\xivt{{\xi_\sst{V}^{(3)}}}
\def\xive{{\xi_\sst{V}^{(8)}}}
\def\xiaz{{\xi_\sst{A}^{(0)}}}
\def\xiat{{\xi_\sst{A}^{(3)}}}
\def\xiae{{\xi_\sst{A}^{(8)}}}
\def\xivtez{{\xi_\sst{V}^{T=0}}}
\def\xivteo{{\xi_\sst{V}^{T=1}}}
\def\xiatez{{\xi_\sst{A}^{T=0}}}
\def\xiateo{{\xi_\sst{A}^{T=1}}}
\def\xiva{{\xi_\sst{V,A}}}

\def\rvz{{R_\sst{V}^{(0)}}}
\def\rvt{{R_\sst{V}^{(3)}}}
\def\rve{{R_\sst{V}^{(8)}}}
\def\raz{{R_\sst{A}^{(0)}}}
\def\rat{{R_\sst{A}^{(3)}}}
\def\rae{{R_\sst{A}^{(8)}}}
\def\rvtez{{R_\sst{V}^{T=0}}}
\def\rvteo{{R_\sst{V}^{T=1}}}
\def\ratez{{R_\sst{A}^{T=0}}}
\def\rateo{{R_\sst{A}^{T=1}}}

\def\mro{{m_\rho}}
\def\mks{{m_\sst{K}^2}}
\def\mpi{{m_\pi}}
\def\mpis{{m_\pi^2}}
\def\mom{{m_\omega}}
\def\mphi{{m_\phi}}
\def\Qhat{{\hat Q}}

\def\FOS{{F_1^{(s)}}}
\def\FTS{{F_2^{(s)}}}
\def\GAS{{G_\sst{A}^{(s)}}}
\def\GES{{G_\sst{E}^{(s)}}}
\def\GMS{{G_\sst{M}^{(s)}}}
\def\GATEZ{{G_\sst{A}^{\sst{T}=0}}}
\def\GATEO{{G_\sst{A}^{\sst{T}=1}}}
\def\mdax{{M_\sst{A}}}
\def\mustr{{\mu_s}}
\def\rsstr{{r^2_s}}
\def\rhos{{\rho_s}}
\def\GEG{{G_\sst{E}^\gamma}}
\def\GEZ{{G_\sst{E}^\sst{Z}}}
\def\GMG{{G_\sst{M}^\gamma}}
\def\GMZ{{G_\sst{M}^\sst{Z}}}
\def\GEn{{G_\sst{E}^n}}
\def\GEp{{G_\sst{E}^p}}
\def\GMn{{G_\sst{M}^n}}
\def\GMp{{G_\sst{M}^p}}
\def\GAp{{G_\sst{A}^p}}
\def\GAn{{G_\sst{A}^n}}
\def\GA{{G_\sst{A}}}
\def\GETEZ{{G_\sst{E}^{\sst{T}=0}}}
\def\GETEO{{G_\sst{E}^{\sst{T}=1}}}
\def\GMTEZ{{G_\sst{M}^{\sst{T}=0}}}
\def\GMTEO{{G_\sst{M}^{\sst{T}=1}}}
\def\lamd{{\lambda_\sst{D}^\sst{V}}}
\def\lamn{{\lambda_n}}
\def\lams{{\lambda_\sst{E}^{(s)}}}
\def\bvz{{\beta_\sst{V}^0}}
\def\bvo{{\beta_\sst{V}^1}}
\def\Gdip{{G_\sst{D}^\sst{V}}}
\def\GdipA{{G_\sst{D}^\sst{A}}}
\def\fks{{F_\sst{K}^{(s)}}}
\def\FIS{{F_i^{(s)}}}
\def\fpi{{f_\pi}}
\def\fk{{f_\sst{K}}}

\def\RAp{{R_\sst{A}^p}}
\def\RAn{{R_\sst{A}^n}}
\def\RVp{{R_\sst{V}^p}}
\def\RVn{{R_\sst{V}^n}}
\def\rva{{R_\sst{V,A}}}
\def\xbb{{x_B}}

\def\mlq{{M_\sst{LQ}}}
\def\mlqs{{M_\sst{LQ}^2}}
\def\lscal{{\lambda_\sst{S}}}
\def\lvect{{\lambda_\sst{V}}}

\def\PR#1{{{\em   Phys. Rev.} {\bf #1} }}
\def\PRC#1{{{\em   Phys. Rev.} {\bf C#1} }}
\def\PRD#1{{{\em   Phys. Rev.} {\bf D#1} }}
\def\PRL#1{{{\em   Phys. Rev. Lett.} {\bf #1} }}
\def\NPA#1{{{\em   Nucl. Phys.} {\bf A#1} }}
\def\NPB#1{{{\em   Nucl. Phys.} {\bf B#1} }}
\def\AoP#1{{{\em   Ann. of Phys.} {\bf #1} }}
\def\PRp#1{{{\em   Phys. Reports} {\bf #1} }}
\def\PLB#1{{{\em   Phys. Lett.} {\bf B#1} }}
\def\ZPA#1{{{\em   Z. f\"ur Phys.} {\bf A#1} }}
\def\ZPC#1{{{\em   Z. f\"ur Phys.} {\bf C#1} }}
\def\etal{{{\em   et al.}}}

\def\delalr{{{delta\alr\over\alr}}}
\def\pbar{{\bar{p}}}
\def\lamchi{{\Lambda_\chi}}

\def\qw0{{Q_\sst{W}^0}}
\def\qwp{{Q_\sst{W}^P}}
\def\qwn{{Q_\sst{W}^N}}
\def\qwe{{Q_\sst{W}^e}}
\def\qem{{Q_\sst{EM}}}

\def\gae{{g_\sst{A}^e}}
\def\gve{{g_\sst{V}^e}}
\def\gvf{{g_\sst{V}^f}}
\def\gaf{{g_\sst{A}^f}}
\def\gvu{{g_\sst{V}^u}}
\def\gau{{g_\sst{A}^u}}
\def\gvd{{g_\sst{V}^d}}
\def\gad{{g_\sst{A}^d}}

\def\gvftil{{\tilde g_\sst{V}^f}}
\def\gaftil{{\tilde g_\sst{A}^f}}
\def\gvetil{{\tilde g_\sst{V}^e}}
\def\gaetil{{\tilde g_\sst{A}^e}}
\def\gvqtil{{\tilde g_\sst{V}^e}}
\def\gaqtil{{\tilde g_\sst{A}^e}}
\def\gvutil{{\tilde g_\sst{V}^e}}
\def\gautil{{\tilde g_\sst{A}^e}}
\def\gvdtil{{\tilde g_\sst{V}^e}}
\def\gadtil{{\tilde g_\sst{A}^e}}

\def\hvf{{h_\sst{V}^f}}
\def\hvu{{h_\sst{V}^u}}
\def\hvd{{h_\sst{V}^d}}
\def\hve{{h_\sst{V}^e}}
\def\hvq{{h_\sst{V}^q}}

\def\delp{{\delta_P}}
\def\delzp{{\delta_{00}}}
\def\deld{{\delta_\Delta}}
\def\dele{{\delta_e}}

\def\apv{{A_\sst{PV}}}
\def\apvnsid{{A_\sst{PV}^\sst{NSID}}}
\def\apvnsd{{A_\sst{PV}^\sst{NSD}}}
\def\qpv{{Q_\sst{W}}}

\def\lamchi{{\Lambda_\chi}}
\def\lamchis{{\Lambda_\chi^2}}
\def\lamchic{{\Lambda_\chi^3}}
\def\FOa{{F_1^{(a)}}}
\def\FTa{{F_2^{(a)}}}
\def\GAS{{G_\sst{A}^{(s)}}}
\def\GEa{{G_\sst{E}^{(a)}}}
\def\GMa{{G_\sst{M}^{(a)}}}
\def\c{{{\cal C}}}
\def\muC{{\tilde{\mu}_{\mbox{{\tiny C}}}}}
\def\muT{{\tilde{\mu}_{\mbox{{\tiny T}}}}}
\def\mus{{\mu_s}}
\def\barq{{\bar q}}
\def\bars{{\bar s}}
\def\bo{{b_0}}
\def\bp{{b_+}}
\def\bm{{b_-}}
\def\bix{{b_9}}
\def\bx{{b_{10}}}
\def\bxi{{b_{11}}}
\def\co{{c_0}}
\def\cp{{c_+}}
\def\cm{{c_-}}
\def\musc{{\tilde{\mu}^s_{\mbox{{\tiny C}}}}}
\title{Baryon Octet magnetic moments in $\chi$PT: More on the importance of
the Decuplet}

\author{S.J. Puglia$^a$ \\
M.J. Ramsey-Musolf$^{a,b}$}
\address{
$^a$ Department of Physics, University of Connecticut,
Storrs, CT 06269 USA \\
$^b$Theory Group, Thomas Jefferson National Laboratory, Newport News,
VA  23606 }


\maketitle
\begin{abstract}

We address the impact of treating the  decuplet of spin-$3\over2$ baryons
as an explicit degree of freedom
in the chiral expansion of the magnetic moments of the octet of
spin-$1\over2$ baryons. We carry out a
complete calculation of the octet moments to ${\cal O}(1/\lamchic)$,
including decuplet contributions to
the chiral loops. In contrast to results of previous analyses, we find that
inclusion of the decuplet preserves
the convergence behavior of the chiral expansion implied by power counting
arguments.

\end{abstract}

\bigskip

The application of Heavy Baryon Chiral Perturbation Theory (HB$\chi$PT) to
low energy baryon properties has
yielded considerable insight.  For example, baryon masses, Compton
scattering amplitudes,  nucleon polarizabilities,
sigma terms, axial couplings,  and hyperon decays have all been
investigated, \cite{eg}. To a large extent, a consistent
description based on chiral symmetry has emerged. The electromagnetic (EM)
properties of baryons have also been studied,
with particular emphasis on the magnetic moments of the lowest-lying baryon
octet (\cite{JMLS},\cite{MS},\cite{DH}).
In contrast to the situation with other baryon properties, the success of
an HB$\chi$PT description of these moments
has been debated. The point of controversy has been whether the chiral
expansion of the octet moments behaves as one
would na\"ively expect, based on power counting arguments. Specifically, if
$\mu_B$ is a generic octet magnetic moment,
one expects its chiral expansion to go schematically as

\begin{equation}
\label{eq:chiexp}
\mu_b= \mu_B^{(0)} +
\mu_B^{(1)}\left({p\over\lamchi}\right)+\mu_B^{(2)}\left({p\over\lamchi}\right)^
2+\cdots\ \ \ ,
\end{equation}
where $\mu_B^{(0)}$ is the tree-level magnetic moment, $p$ is of order the
pseudoscalar meson masses, $\lamchi=4\pi
f_\pi\approx 1$ GeV, and the $\mu_B^{(n)}$, $n>1$, represent the
long-distance (loop) and short-distance (counterterm)
corrections to $\mu_B^{(0)}$ at a given order. To the extent that the
$\mu_B^{(n)}$ are all of a similar order of
magnitude, the relative size of successive terms in the expansion of Eq.
(\ref{eq:chiexp}) should decrease by the
corresponding power of $(p/\lamchi)$. Taking $p\sim m_K$, for example, the
expansion parameter should be or order
$m_K/\lamchi\sim 1/2$.

The degree to which this behavior holds for the magnetic moments has been
debated, and various remedies have been
proposed for the apparent deviation of the expansion from this expectation.
These remedies include adjusting the size
of tree-level axial couplings to reduce the scale of kaon loop
contributions \cite{JMLS}, inclusion of $1/\mn$ corrections
and higher-derivative terms \cite{MS}, explicit retention of the leading
analytic (in quark mass) loop contributions
\cite{MS}, and inclusion of the decuplet \cite{DH}. To date, no ${\cal
O}(q^4)$ calculation has included both octet
and decuplet loop contributions as well as the leading $1/\mn$ corrections
and two-derivative operators. In an attempt
to resolve some of the controversy, we have performed such a calculation.
We find that that when decuplet is included as
an explicit degree of freedom -- along with the leading $1/\mn$ corrections
and two-derivative contributions -- the
magnetic moment expansion behaves according to na\"ive power counting
expectations (Eq. (\ref{eq:chiexp})) {\em without}
including analytic loop contributions or adjusting the size of tree-level
axial couplings.

Before discussing our calculation in detail, we review the history of
HB$\chi$PT magnetic moment analyses.
Jenkins and Manohar \cite{JM1} note that due to the
strength of the decuplet-octet coupling $\c$, and the  small size of the
mass splitting $\delta$ (relative to the intrinsic hadronic scale), the
contributions from the
decuplet should be larger than those from higher baryon resonances and
comparable to the octet contribution.
In the case of the axial current \cite{JM2} substantial cancellations
occur. It is then surprising that in
the ${\cal O}(q^4)$  magnetic moment calculation of Ref. \cite{JMLS},
inclusion of the decuplet
does not produce appreciably better agreement with the data than the case
where only the octet was included. The authors
of Ref. \cite{JMLS} argue that there is some evidence that $\chi$PT
overestimates the size of the kaon
loops. They propose compensating for this effect by using the smaller
one-loop corrected axial couplings in the
calculation rather than the tree-level values.

The study of Ref. \cite{MS}, however,
suggests that the calculation of \cite{JMLS} is incomplete. In particular,
it neglects the $1/\mn$ corrections and the
contributions of certain double derivative operators, both of which occur
at ${\cal O}(q^4)$.
The decuplet was not included explicitly in the analysis of Ref.\cite{MS}. Its
effect was only considered in determining its contribution to some low
energy constants (LEC). These  LEC's appear as
couplings to operators of the form $\partial_\mu\phi\partial_\nu\phi$,
where $\phi$ denotes the Goldstone boson field, which
contribute at ${\cal O}(q^4)$. The same LEC's also receive contributions
from the vector mesons and higher lying resonances
such as the Roper octet. By expanding decuplet loop amplitudes in powers of
$1/\delta$, the authors of
Ref.\cite{MS} argue that the LEC's in question contain the leading order
decuplet contribution.  They conclude that there
is no need to employ smaller values for the axial couplings or to include
the decuplet explicitly. They also conclude from
their calculation that the chiral expansion converges as expected. We note,
however, that Ref. \cite{MS} retains
contributions analytic in the quark mass which arise from loop
contributions at ${\cal O}(q^4)$.

The later analysis of Durand and Ha {\cite{DH} re-examines the analysis of
Ref.\cite{JMLS}, and focuses mainly on the
convergence of  the expansion. These authors argue that the decuplet must
be included explicitly. They observe
that the treatment of Ref.\cite{MS} requires  the decuplet-octet mass
splitting to
be large with  respect to the momentum in the loop integrals, which is not
the case. They conclude that since the
mass splittng is approximately 300  MeV, the decuplet must be considered as
\lq\lq light" and included explicitly.
A similar observation appears in the work of Banerjee {\em et
al.}\cite{Ban}. The authors of Ref. \cite{DH} conclude that
the chiral expansion for the magnetic moments of the octet is not
convergent.  However, they did not include all of the
terms contributing at ${\cal O}(q^4)$.

In what follows, we attempt to resolve this controversy.
We do so by carrying out the a complete analysis of the octet magnetic
moment at ${\cal
O}(q^4)$ by including the decuplet explicitly and the full set of $1/\mn$
corrections. Our analysis is similar in spirit
to that of Ref.\cite{MS}, but differs in two respects: (a) the explicit
inclusion of the decuplet, and (b) retention of only
non-analytic loop contributions.

To make our notation and conventions clear, we review some elements of the
HB$\chi$PT formalism.
In this formalism a consistent chiral expansion of the baryon Lagrangian
can be written in terms of
the
velocity-dependent octet and decuplet fields:
\begin{eqnarray}
  B_v(x)=\exp(i M_B{\not\! v}v\cdot x) B(x);&\  T^\mu_{v}(x)=
  \exp(i M_B{\not\! v}v\cdot x)T^\mu(x)&\ \ ,
\end{eqnarray}
here $B(x)$ and $T^\mu(x)$ denote the baryon octet and decuplet fields,
respectively, and
$M_B$ is the SU(3) invariant mass of the octet. Defining the fields in this
way eliminates
ambiguities in power counting that arise due to the introduction of another
large mass scale in the
theory, {\em i.e.} the octet mass \cite{JM1}.

The leading order contributions to the magnetic moments
of the octet are calculated from tree level graphs with vertices from the
Lagrangians (we use the
notation of \cite{MRM/ITO}):
\begin{eqnarray}
{\cal L}_{1}&=& \frac{e}{\lamchi}\ \epsilon_{\mu\nu\rho\sigma} v^\rho
\left\{b_{+}\hbox{ Tr}\left({\bar{B}}_v S^{\sigma}_v \{Q,B_b\}\right)
 +b_{-}\hbox{ Tr}\left({\bar{B}}_v S^{\sigma}_v
[Q,B_v]\right)\right\}F^{\mu\nu} \label{eq:L1}
\end{eqnarray}
Here we have introduced the covariant spin operator $S^{\mu}_v$
whose properties are discussed in Ref.\cite{JM1}. We choose to normalize in
powers of ${1/ \lamchi}\
(\lamchi= 4\pi\fpi\approx 1$GeV). In what follows, we count in powers of
$1/\lamchi$ rather than in powers of $q$ as is done in Ref.\cite{MS} (for
instance, the conversion
${\cal O}(q^4)\leftrightarrow{\cal O}(1/\lamchic) $applies).

One-loop corrections are generated using the vertices from the
lowest order chiral Lagrangian for octet and decuplet baryons which depends
on the octet of
pseudoscalar mesons $\tilde{\Pi}$. Introducing the the non-linear
representation of the mesons
 \begin{eqnarray}
 \xi&=&e^{i\tilde{\Pi}/\fpi}\\
 \Sigma&=&\xi^2
 \end{eqnarray}
 and defining the vector and axial vector combinations

 \begin{eqnarray}
V_\mu & \equiv & {1\over 2}(\xi^{\dag}\partial_\mu\xi+\xi\partial_\mu
	\xi^{\dag})
\label{eq:vec}\\
	A_\mu & \equiv & {i\over 2}(\xi^{\dag}\partial_\mu\xi-\xi\partial_\mu
	\xi^{\dag})\ \ \ ,
\label{eq:axi}
\end{eqnarray}
we write the lowest order Lagrangian (using the notation of \cite{JMLS})
\begin{eqnarray}
{\cal L}_0&=&i\hbox{ Tr}\left({\bar B}_vv\cdot D\ B_v\right)+ 2D
\hbox{ Tr}\left({\bar B}_v S_v^\mu\{A_\mu, B_v\}\right)\nonumber \\
& &+2F\hbox{ Tr}
\left({\bar B}_v S_v^\mu[A_\mu, B_v]\right)  -i{\bar T}^\mu_v \left(v\cdot
{\cal D}\right)T_{v\mu} \nonumber \\
& &+\delta {\bar T}^\mu_v T_{v\mu}+\c\left({\bar T}^\mu_v A_\mu B_v+{\bar
B}_v A_\mu T_{v\mu}\right)+ 2{\cal H}{\bar
T}^\mu_v S_v^\nu A_\nu T_{v\mu}+{\fpi^2\over
4}\hbox{Tr}\left(\partial^\mu\Sigma^{\dag}
   \partial_\mu\Sigma\right).
\label{eq:lheav}
\end{eqnarray}
Here $\delta=M_{\sst{T}}-M_{\sst{B}}$ is the baryon octet-decuplet mass
splitting which arises due to the way we defined the velocity dependent
fields. We use $D=.75$,
$F=.50$
and ${\cal C}=-1.5$ throughout \cite{MS}. The value of ${\cal H}$ is
not needed here.
Interactions due to the vector current $V_\mu$ appear in the chiral
covariant derivatives
\[
D_\mu B= \partial_\mu B +[V_\mu,B]
\]
and
\[
{\cal D}_\nu T^\mu_{ijk}= \partial_\nu T^\mu_{ijk} +(V_\nu)^l_i
T^\mu_{ljk}+(V_\nu)^l_j T^\mu_{ilk}+(V_\nu)^l_k T^\mu_{ijl}
\]
where $i,j,k=1,2,3$ are SU(3) flavor indices.

The electromagnetic interaction is incorporated into ${\cal L}_0$ via the
substitutions
\begin{eqnarray}
V_\mu&\rightarrow& V_\mu+{1\over 2} i e {\cal A}_\mu\left(\xi^\dag Q\xi
+\xi Q\xi^\dag\right) \\
A_\mu&\rightarrow& A_\mu-{1\over 2}  e {\cal A}_\mu\left(\xi^\dag Q\xi -\xi
Q\xi^\dag\right)
\end{eqnarray}
and
\begin{eqnarray}
\partial_\mu\Sigma&\rightarrow& \partial_\mu\Sigma+{1\over 2} i e {\cal
A}_\mu[Q,\Sigma]\ ,
\end{eqnarray}
where ${\cal A}_\mu$ is the photon field. The full chiral structure of the
Lagrangian
${\cal L}_1$ is given by the replacement
\begin{eqnarray}
Q&\rightarrow& {1\over 2}\left(\xi^\dag Q\xi +\xi Q\xi^\dag\right)\ \ \ .
\end{eqnarray}
 For the magnetic moments to ${\cal O}(1/\lamchic)$ at one-loop there are
further
contributions. First there are insertions of the leading order moments into
the loops
as well as insertions of the decuplet magnetic moment and the
octet-decuplet transition moments.
The decuplet magnetic moment operator can be written
\begin{equation}
{\cal L}_{\mbox{{\tiny T}}}= -ie \ \frac{\tilde{\mu}_{\mbox{{\small
c}}}q_i}{\lamchi}{\bar T}^\mu_{vi}
T^{\nu}_{vi}F_{\mu\nu}\ ,
\end{equation}
where $q_i$ is the charge of the $i$th member of the decuplet.
The measured value of the $\Omega^{-}$ moment determines
$\tilde{\mu}_{\mbox{c}}=
1.20\pm 0.14$ (in our normalization). The octet-decuplet transition operator
 is given by (\cite{JMLS} and references therein)
 \begin{equation}
 {\cal L}_{\mbox{{\tiny BT}}}=ie\  \frac{\muT}{\lamchi}\left(
 \epsilon_{ijk}Q^i_l \bar{B}^j_m S^\mu_v T^{\nu klm}+\epsilon^{ijk}Q^l_i
  \bar{T}^\mu_{lkm} S^\nu_v B^m_j\right)F_{\mu\nu}\ ,
  \end{equation}
where $i,j,k,l,m= 1,2,3$ are flavor indices.
Measured values for $\Delta\rightarrow\gamma N$ helicity amplitudes determine
$\muT= -4.79\pm0.31$ (again in our normalization).

There is an additional set of dimension five operators which generates the
double derivative operators mentioned above.
They contribute to loops at ${\cal O}(1/\lamchic)$ \cite{MS} and
are given by

\begin{eqnarray}
{\cal L}_{\mbox{\tiny MB}}&=&\frac{4i}{\lamchi} \
\epsilon_{\mu\nu\rho\sigma} v^\rho
\left\{b_9\hbox{ Tr}({\bar{B}}_v S^{\sigma}_v A^\mu)\hbox{ Tr}\left(A^\nu
B_v\right)+
b_{10}\hbox{ Tr}\left({\bar{B}}_v S^{\sigma}_v
\left[A^\mu,A^\nu\right]B_v\right)\right. \nonumber\\
& &\left.+b_{11}\hbox{ Tr}\left({\bar{B}}_v S^{\sigma}_v
\{A^\mu,A^\nu\}B_v\right)\right\}.
\label{eq:dbld}
\end{eqnarray}

The loops derived from the operators listed above generate 
${\cal O}(1/\lamchis)$ and ${\cal O}(1/\lamchic)$
contributions. Additional contributions of ${\cal O}(1/\lamchis\mn)$ are
obtained from the $1/\mn$ expansion of the lowest
order Lagrangian. Only the corrections to the baryon propagators contribute
and are given by \cite{Hem}

\begin{eqnarray}
{\cal L}_{1\over\mn}&=& \frac{1}{2\mn}\left\{\hbox{
Tr}\left({\bar{B}}_v[v\cdot D,[v\cdot
D,B_v]]\right)
-\hbox{ Tr}\left({\bar{B}}_v[ D^\mu,[D_\mu,B_v]]\right)\right.\nonumber\\
&+&\left.{\bar T}^\mu_v\left({\cal D}^\alpha {\cal D}_\alpha-v\cdot {\cal
D}\ v\cdot {\cal
D}\right)T_{v\mu}\right\}.
\end{eqnarray}

Finally, along with the above couplings, the calculation of the magnetic
moments
requires the introduction of counter terms which break chiral SU(3)
symmetry at
${\cal O}(1/\lamchic)$.
\begin{eqnarray}
{\cal L}_{\mbox{\tiny SB}}&=& \frac{e}{\lamchi}\
\epsilon_{\mu\nu\rho\sigma} v^\rho
F^{\mu\nu}\left\{
b_3 \hbox{ Tr}\left(\bar{B}_v
S^{\sigma}_v\left[[Q,B],{\cal M}\right]\right)
+b_4\hbox{ Tr}\left(\bar{B}_v
S^{\sigma}_v\left\{[Q,B],{\cal M}\right\}\right)\right.\nonumber \\
& &\left.+b_5 \hbox{ Tr}\left(\bar{B}_v S^{\sigma}_v\left[\{Q,B\},{\cal
M}\right]\right)
+b_5 \hbox{ Tr}\left(\bar{B}_v
S^{\sigma}_v\left\{\{Q,B\},{\cal M}\right\}\right)
+b_7\hbox{ Tr}\left(\bar{B}_v S^{\sigma}_vB\right)\hbox{ Tr}\left({\cal
M}Q\right)\right\}.
\label{eq:Lsb}
\end{eqnarray}
Chiral SU(3)-breaking is introduced through the strange quark mass by the
matrix {${\cal M}= B_0 m_s\mbox{diag}(0,0,1)$},
where $B_0$ carries dimensions of mass and is related to the scalar quark
condensate. Since ${\cal M}$
counts as two powers of the meson mass it appears that
our normalization for these symmetry breaking terms is incorrect. This is
not the case. It might seem more natural to write the coupling and
normalization as
\[
 {B_0 m_s\tilde{b}_i \over \lamchic} \hspace{1cm} (i= 3-7)
 \]
 with the $\tilde{b}_i$'s of ${\cal O}(1)$. However, we avoid taking
 explicit values for $B_0$ and $m_s$ by absorbing these and
 two powers of $\lamchi$ into our couplings, thus writing
 \[
 b_i = {B_0 m_s\tilde{b}_i \over \lamchis}
 \]
so that our normalization follows. With this choice, all the tree graphs
appearing in the calculation
of the magnetic moment appear to be of ${\cal O}(1/\lamchi)$. However, they
contribute at different
orders and now the couplings for the symmetry breaking terms are no longer
of natural size.

Using the above conventions, we compute the EM magnetic moments to
${\cal O}(1/\lamchic)$ as generated by the tree-level
operators of Eqs.(\ref{eq:L1}, \ref{eq:Lsb}) and the non-analytic
contributions from the one-loop
graphs of Fig. 1. Following a similar notation to that of \cite{DH} we
write the results for
the magnetic moments as

\begin{eqnarray}
\lefteqn{\mu_B=
\left(\frac{2M_N}{\lamchi}\right)\left\{\alpha_B+ \frac{\pi}{\lamchi}
\sum_{X=\pi,K}\left(\beta_B^{(X)} m_X+\beta^{\prime (X)}_B
F(m_X,\delta,\mu)\right)\right.}\nonumber \\
\ &\ &\hspace*{1cm}\left.+\frac{1}{\lamchis}\sum_{X=\pi,K,\eta}\left[
\left(\gamma^{(X)}_B-\lambda^{(X)}_B\alpha_B
+{5\over2\mn}(\beta_B^{(X)}+{1\over6}\beta^{\prime (X)}_B)
\right)m_X^2\ln\frac{m_X^2}{\mu^2}\right.\right.\nonumber\\
\ &\ &\hspace*{6cm}\left.
\left.+\left(\tilde{\gamma}^{(X)}_B-\tilde{\lambda}^{(X)}_B\alpha_B\right)L_{(3/
2)}
+\hat{\gamma}^{(X)}_B\hat{L}_{(3/2)}\right]\right\},
\label{eq:mu1}
\end{eqnarray}
where
\begin{equation}
\begin{array}{l}

\pi F(m,\delta,\mu)=-\delta\ln{m^2\over\mu^2}+\left\{\begin{array}{ll}
2\sqrt{m^2-\delta^2}\left({\pi\over 2}-
\arctan\left[{\delta\over\sqrt{m^2-\delta^2}}\right]\right)&
 m\geq \delta\nonumber \\
  & \nonumber \\
 -2\sqrt{\delta^2-m^2}\ln\left[{\delta+\sqrt{\delta^2-m^2}\over m}\right]&
 m<\delta
 \end{array}
 \right.
  \\ \\
 L_{(3/2)}(m,\delta,\mu)=m^2\ln{m^2\over\mu^2}+2\pi\delta F(m,\delta,\mu)
\\ \\
\hat{L}_{(3/2)}(m,\delta,\mu)=m^2\ln{m^2\over\mu^2}+{2\pi\over3\delta}
G(m,\delta,\mu)
\\ \\

\pi G(m,\delta,\mu)=-\delta^3\ln{m^2\over\mu^2}+\pi m^3-
\left\{\begin{array}{ll}
2(m^2-\delta^2)^{3/2}\left({\pi\over2}-
\arctan\left[{\delta\over\sqrt{m^2-\delta^2}}\right]\right)&
 m\geq \delta\nonumber \\
  & \nonumber \\
 2(\delta^2-m^2)^{3/2}\ln\left[{\delta+\sqrt{\delta^2-m^2}\over m}\right]&
 m<\delta .
 \end{array}
 \right.
 \end{array}\label{eq:defs}
 \end{equation}
Here $\mu$ is the scale of dimensional regularization whose value we take
as 1 GeV in the following.
 The coefficients $\alpha_B$ are the tree level contributions
 which are linear combinations of the
 coupling constants appearing Eq. (\ref{eq:L1}) and Eq. (\ref{eq:Lsb});
  $\beta^{(X)}_B$ and $\beta^{\prime(X)}_B$ are the contributions
  from the meson one-loop graphs in
 Figs 1a. containing intermediate octet and decuplet states, respectively;
 $\gamma^{(X)}_B$,$\tilde{\gamma}^{(X)}_B$
 and $\hat{\gamma}^{(X)}_B$ are the contributions from the graphs in Fig. 1
b,c,d, respectively; $\lambda^{(X)}_B$ and
 $\tilde{\lambda}^{(X)}_B$ are the wavefunction renormalization
contributions, again with octet and decuplet
intermediate states, respectively.
Due to our choice of normalizations, our coefficients appear different from
those given in \cite{JMLS},\cite{MS} and \cite{DH} so we list them all in
Appendix I. They are, however, in complete agreement with those references
including the corrections noted in \cite{DH} and the Erratum to \cite{JMLS}.

 We turn now to a determination of the low-energy constants $b_{\pm}$,
 and $b_i$, $i=3,\ldots, 7$. It is instructive to consider the evolution of
these constants as the chiral expansion is carried out to successively
higher orders and as
decuplet intermediate states are included. For the magnetic moments at ${\cal
O}(1/\lamchi)$ and ${\cal O}(1/\lamchis)$, we perform an un-weighted least
squares fit of the
leading order constants $b_{\pm}$ using the seven well-measured octet
magnetic moments
(we exlude the $\mu_{\Sigma^0}$). At ${\cal O}(1/\lamchic)$, there appear
the five
additional symmetry breaking constants plus the unknown constants
$b_9-b_{11}$. We follow the same
procedure as Ref.\cite{MS} by using resonance saturation to determine
$b_9-b_{11}$. The details can be
found in that reference. We note, however, that where we have taken the
decuplet as an explicit degree
of freedom we do not include its contribution to these couplings. We chose
to leave the symmetry
breaking constants as fit parameters and use the seven well-measured
moments to obtain an exact
solution. The values of the $b_i$ are given in Table I, for
two scenarios: (O) -- only the octet loop corrections included at a given
order, and (O+D)
-- both octet and decuplet loop effects included. A measure of the quality
of the fits is given in
Table II, where the magnetic moments predicted at a given order are
compared with the experimental
values (final two columns). At ${\cal O}(1/\lamchic)$ only the
$\Sigma^0-\Lambda$ transition moment
is a prediction.

As observed in previous analyses, at ${\cal O}(1/\lamchis)$ the fit without
the decuplet
($\chi^2=0.377$) is better than the one where it is included
($\chi^2=0.651$). Evidently, truncation
of the chiral expansion at ${\cal O}(1/\lamchis)$ is not sufficient in this
case. At ${\cal
O}(1/\lamchic)$, the fits are exact and it is difficult to see the effect
of the decuplet. To gain
some insight we examine the contribution from each order individually. As
an example we consider the
magnetic moment of the proton. In the case where only intermediate octet
states are considered the
magnetic moment breaks down as follows

\begin{equation}
\mu_p=3.268(1-.687+.541)=2.791. \label{eq:Mu1}
\end{equation}
Here we have normalized to the tree level moment, which arises at ${\cal
O}(1/\lamchi)$. The second and third
terms in parentheses correspond to the ${\cal O}(1/\lamchis)$ and ${\cal
O}(1/\lamchic)$ corrections, respectively.
We see that the contribution from the ${\cal O}(1/\lamchic)$ terms are as
large as those contributing at ${\cal
O}(1/\lamchis)$.  In the case where the decuplet is included
we obtain

\begin{equation}
\mu_p=4.695(1-.513+.108)=2.793. \label{eq:Mu2}
\end{equation}
Here we see the effect of the decuplet. Na\"\i vely, one would expect the
corrections to scale as $(p/\lamchi)$ and
$(p/\lamchi)^2$, respectively, relative to the  tree-level contribution.
Taking $p=m_k$, then one expects the various orders
to contribute as $1:1/2:1/4$ or as $1:1/3:1/9$ using $p=\delta$. Clearly, this 
pattern does
not obtain in the case of the octet only calculation, but does in the
octet+decuplet case. We find a similar conclusion for
each octet magnetic moment.
\[
\begin{array}{ccccc}
\mu_p&=&4.695(1-.513+.103)&=&2.793 \\
\mu_n&=&-3.204(1-.446+.043)&=&-1.913 \\
\mu_{\Xi^-}&=&-1.491(1-.703+.141)&=&-0.653 \\
\mu_{\Xi^0}&=&-3.204(1-.929+.319)&=&-1.250 \\
\mu_{\Sigma^+}&=&4.695(1-.707+.230)&=&2.450 \\
\mu_{\Sigma^-}&=&-1.491(1-.184-.039)&=&-1.160 \\
\mu_\Lambda&=&-1.601(1-.950+.332)&=&-0.613 \\
\mu_{\Sigma^0\Lambda}&=&2.775(1-.600+.138)&=&-1.491 
\end{array}
\]
Apart from a few exceptions ($\mu_\Lambda,\ \mu_{\Xi^0}$ at ${\cal 
O}(1/\lamchis)$, the chiral expansion
seems to converge as expected when the
${\cal O}(1/M_{\mbox{\tiny B}})$ corrections and the decuplet are included
explicitly. The result in Eq.(\ref{eq:Mu2}) should be compared to that of 
Ref.\cite{MS}

\begin{equation}
\mu_p=4.48(1-.49+.11)=2.79. \label{eq:Mu3}
\end{equation}

Eqs.(\ref{eq:Mu2}) and (\ref{eq:Mu3}) are essentially identical, so we must
make some comment on how they differ.
Eq.(\ref{eq:Mu1}) is the result of a calculation similar to the one made in
Ref.\cite{MS} to obtain Eq.(\ref{eq:Mu3}).
They are different only in that we include only the non-analytic
contributions from loops. In Ref. \cite{MS} some of the loops appearing at 
${\cal O}(1/\lamchic)$ have the analytic structure
\[
\mbox{constant}\times (m_X^2\ln{m_X^2\over\mu^2} - m_X^2)
\]
where $X=\pi, K$.
Evidently, the inclusion of the analytic piece cancels a significant
portion of the non-analytic one. To be
specific, for the pions the cancellation is about 25\% and for kaons it is
greater than 70\%. Since the analytic piece of a
loop (or any portion of it) can be absorbed into the counterterms, the
prescription for retaining it explicitly is
ambiguous. It is satisfying to see that we can obtain the expected behavior
of the expansion without resorting to this
procedure.

To exhibit the importance of including the $1/\mn$ corrections and the
contribution from the double derivative
terms  we compare Eq.(\ref{eq:Mu2}) to the result of Ref.\cite{DH}
\begin{equation}
\mu_p=3.668(1-.651+.412)=2.791, \label{eq:Mu4}
\end{equation}
Here again the contribution from the ${\cal O}(1/\lamchic)$ terms is a
larger fraction of those from
${\cal O}(1/\lamchis)$ than expected.

In summary we have re-examined the calculation of the magnetic moments for
the octet of spin-$1\over2$ baryons to
${\cal O}(1/\lamchic)$ in HB$\chi$PT. We have included all terms which
contribute to this order.
The decuplet of spin-$3\over2$ was included as an explicit degree of
freedom and its contribution to the octet magnetic
moments evaluated. Our analysis indicates that including the
decuplet is necessary to insure the correct size of the the contributions
from succeeding orders in the chiral expansion.
Only the non-analytic contributions of loops were retained so we avoid the
ambiguities involved in including any analytic
pieces. We also find no need to take smaller values for the axial
couplings. Thus, it appears that a well-behaved,
consistent chiral expansion of the octet baryon magnetic moments is
attainable at ${\cal O}(q^4)$.\footnote{In this respect we also mention here the 
analysis of Ref.\cite{Hol}. The authors of that Ref.\cite{Hol}
develop a regularization scheme in which a momentum cut-off is introduced to 
suppress short-distance contributions to the
Feymann integrals. It appears that this procedure may improve the convergence of 
the chiral expansion for the magnetic moments while respecting the chiral 
symmetry.}
This result should put the baryon magnetic moments on the same chiral footing as 
other
low-energy baryon properties. 

\acknowledgements
We would like to thank Martin Savage and Thomas Hemmert for useful discussions. 
We  would also like to thank the Institute for Nuclear Theory for their 
hospitality. This work was supported in part under U.S. Department of Energy 
contract No. DE-FG06-90ER40561 and DE-AC05-84ER40150 and a National Science 
Foundation Young Investigator Award.

\begin{table}
 \begin{tabular}{|c||c|c|c|c|c|}\hline
  & ${\cal O}(1/\lamchi)$&\multicolumn{2}{c |}{${\cal O}(1/\lamchis)$}&
\multicolumn{2}{c |}{${\cal O}(1/\lamchic)$}\\ \hline
 $CT$ & & O    &O+D&O    &O+D\\ \hline
 & & & & &      \\
  $b_{+}$& $1.490$ & $2.999$ & $3.606$ & $1.994$ & $2.989$ \\
  $b_{-}$& $1.098$ & $2.194$ & $2.194$ & $1.368$ & $1.924$ \\
  $b_3$& $-$ & $-$ & $-$ & $-0.297$ & $-0.327$ \\
  $b_4$& $-$ & $-$ & $-$ & $0.277$ & $-0.159$ \\
  $b_5$& $-$ & $-$ & $-$ & $-0.086$ & $0.124$ \\
  $b_6$& $-$ & $-$ & $-$ & $0.576$ & $0.171$ \\
  $b_7$& $-$ & $-$ & $-$ & $-0.952$ & $-1.166$ \\
  \hline
  \end{tabular} 
 
  \caption{ Couplings for leading order and symmetry breaking
magnetic moment counterterms at each order, with and without including the
decuplet intermediate states in loops.
``O" and ``D" denote octet and decuplet respectively.}

\begin{tabular}{|c||c|c|c|c|c|}\hline
  & ${\cal O}(1/\lamchi)$&\multicolumn{2}{c |}{${\cal O}(1/\lamchis)$}&
\multicolumn{2}{c |}{${\cal O}(1/\lamchic)$}\\ \hline
 $\mu_{\mbox{Fit}}$ & & O    &O+D&O    &O+D\\ \hline
 & & & & &      \\
  $p$& $2.564$ & $2.890$ & $3.051$ & $2.793$ & $2.793$ \\
  $n$& $-1.597$ & $-2.360$ & $-2.437$ & $-1.913$ & $-1.913$ \\
  $\Xi^{-}$& $-0.967$ & $-0.585$ & $-0.547$ & $-0.651$ & $-0.651$ \\
  $\Xi^{0}$& $-1.597$ & $-0.933$ & $-0.887$ & $-1.250$ & $-1.250$ \\
  $\Sigma^{+}$& $2.564$ & $2.287$ & $2.141$ & $2.458$ & $2.458$ \\
  $\Sigma^{-}$& $-0.967$ & $-1.298$ & $-1.321$ & $-1.160$ & $-1.160$ \\
  $\Lambda$& $-0.799$ & $-0.494$ & $-0.410$ & $-0.613$ & $-0.613$ \\
  $\Sigma^0\Lambda$& $1.383$ & $1.617$ & $1.6827$ & $1.520$ & $1.491$ \\
\hline
\end{tabular}
\caption{Calculated values of the magnetic moments using fit
values of the counterm couplings.``O" and ``D" denote octet and decuplet
respectively.}
\end{table}
\newpage
  \begin{center}
  {\bf\Large Appendix I.}
  \end{center}
 Here we tabulate the coeffcients appearing the expressions for the
magnetic moments.\\

\begin{center}
\begin{tabular}{|l|l|}\hline
&\multicolumn{1}{|c|}{$\alpha_B$}\\[1mm] \hline
$p$&${1\over3}\bp+\bm+b_3+b_4+{1\over3}b_5+{1\over3}b_6-{1\over3}b_7$\\[2mm]
$n$&$-{2\over3}\bp-{2\over3}b_5-{2\over3}b_6-{2\over3}b_7$\\[2mm]
$\Xi^-$&$-{1\over3}\bp-\bm+b_3-b_4-{1\over3}b_5+{1\over3}b_6-{1\over3}b_7$\\[2mm]
$\Xi^0$&${-2\over3}\bp+{1\over3}b_5-{1\over3}b_6-{1\over3}b_7$\\[2mm]
$\Sigma^+$&${1\over3}\bp+1\bm-{1\over3}b_7$\\[2mm]
$\Sigma^-$&${1\over3}\bp-1\bm-{1\over3}b_7$\\[2mm]
$\Lambda$&$-{1\over3}\bp-{8\over9}b_6-{1\over3}b_7$\\[2mm]
$\Sigma^0\Lambda$&${1\over\sqrt{3}}\bp$\\[2mm]
\hline
\end{tabular}
\end{center}
\vspace*{3cm}
\begin{center}
\begin{tabular}{|l|c|c|}\hline
&\multicolumn{2}{|c|}{$\beta^{(X)}_B$}\\[1mm]  \hline
&$\pi$&$K$\\[1mm] \hline
$p$&$-(D+F)^2$ &$-({2\over3}D^2+2F^2)$ \\[2mm]
$n$&$(D+F)^2$ &$-(D-F)^2$ \\[2mm]
$\Xi^-$&$(D-F)^2$ &$({2\over3}D^2+2F^2)$ \\[2mm]
$\Xi^0$
& $-(D-F)^2$& $(D+F)^2$\\[2mm]
$\Sigma^+$& $-({2\over3}D^2+2F^2)$&$-(D+F)^2$ \\[2mm]
$\Sigma^-$& $({2\over3}D^2+2F^2)$&$(D-F)^2$ \\[2mm]
$\Lambda$& $0$&$2DF$ \\[2mm]
$\Sigma^0\Lambda$&$-{4\over\sqrt{3}}DF$ &$-{2\over\sqrt{3}}DF$ \\[2mm]
\hline
\end{tabular}
\end{center}

\begin{center}
\begin{tabular}{|l|c|c|}\hline
&\multicolumn{2}{|c|}{$\beta^{\prime(X)}_B$}\\[1mm]  \hline
&$\pi$&$K$\\[1mm] \hline
$p$&$-{2\over9}\c^2$ &${1\over18}\c^2$ \\[2mm]
$n$&${1\over18}\c$ &${1\over9}\c^2$ \\[2mm]
$\Xi^-$&$-{1\over9}\c^2$ &$-{1\over18}\c^2$ \\[2mm]
$\Xi^0$
& ${1\over9}\c^2$& ${2\over9}\c^2$\\[2mm]
$\Sigma^+$& ${1\over18}\c^2$&$-{2\over9}\c^2$ \\[2mm]
$\Sigma^-$& $-{1\over18}\c^2$&$-{1\over9}\c^2$ \\[2mm]
$\Lambda$& $0$&${1\over6}\c^2$ \\[2mm]
$\Sigma^0\Lambda$&$-{1\over3\sqrt{3}}\c^2$ &$-{1\over6\sqrt{3}}\c^2$ \\[2mm]
\hline
\end{tabular}
\end{center}

\begin{center}
\begin{tabular}{|l|c|}\hline
&$\gamma^{(\pi)}_B$\\[1mm]  \hline
$p$&$-(\bp+\bm)+{1\over2}(D+F)^2(\bp-\bm)+2(\bx+\bxi)$\\[2mm]
$n$&$\bp+\left(1-(D+F)^2\right)\bm-2(\bx+\bxi)$\\[2mm]
$\Xi^-$&$\bm-\bp+{1\over2}(D-F)^2(\bp+\bm)+2(\bx-\bxi)$\\[2mm]
$\Xi^0$&$\bp-\left(1-(D-F)^2\right)\bm-2(\bx+\bxi)$\\[2mm]
$\Sigma^+$&${2\over9}(D^2+6DF-6F^2)\bp-2(1+F^2)\bm+2(\bx+\bxi)$\\[2mm]
$\Sigma^-$&${2\over9}\left(D^2-6DF-6F^2\right)+2(1+F^2)\bm-\bix-4\bxi$\\[2mm]
$\Lambda$&$-{2\over3}D^2\bp$\\[2mm]
$\Sigma^0\Lambda$&$-{2\over3\sqrt{3}}(3-D^2)\bp+{4\over3\sqrt{3}}DF\bm+{4\over 
\sqrt{3}}\bx$\\[2mm]
\hline
\end{tabular}
\end{center}
\begin{center}
\begin{tabular}{|l|c|}\hline
&$\gamma^{(K)}_B$\\[1mm]  \hline
$p$&$-\left({1\over9} D^2-2DF+F^2\right)\bp-\left(2+(D-F)^2\right)\bm+\bix+4\bxi$\\[2mm]
$n$&$\left(1-{7\over9}D^2+{2\over3}DF+F^2\right)\bp-\left(1-(D-F)^2\right)\bm+2(\bxi-\bx)$\\[2mm]
$\Xi^-$&$-\left({1\over9}D^2+2DF+F^2\right)\bp+\left(1+(D+F)^2\right)\bm-\bix-4\bxi$\\[2mm]
$\Xi^0$&$\left(1-({7\over9}D^2+{2\over3}DF-F^2)\right)\bp+\left(1-(D+F)^2\right)-2(\bx+\bxi)$\\[2mm]
$\Sigma^+$&$\left(({1\over3}D^2+2DF+{1\over3}F^2)-1\right)\bp-\left(1+(D-F)^2\right)\bm+2(\bx+\bxi)$\\[2mm]
$\Sigma^-$&$\left(({1\over3}D^2-2DF+{1\over3}F^2)-1\right)\bp-\left(1+(D+F)^2\right)\bm+2(\bx-\bxi)$\\[2mm]
$\Lambda$&$\left(1+{1\over9}D^2+F^2\right)\bp-2DF\bm-2\bx$\\[2mm]
$\Sigma^0\Lambda$&${1\over\sqrt{3}}\left(D^2-3F^2-1\right)\bp+{2\over\sqrt{3}}DF\bm+{4\over\sqrt{3}}\bx$\\[2mm]
\hline
\end{tabular}
\end{center}

\begin{center}
\begin{tabular}{|l|c|}\hline
&$\gamma^{(\eta)}_B$\\[1mm]  \hline
$p$&$-{1\over18}(D-3F)^2(\bp+3\bm)$\\[2mm]
$n$&${1\over9}(D-3F)^2\bp$\\[2mm]
$\Xi^-$&${1\over18}(D+3F)^2(\bp-3\bm)$\\[2mm]
$\Xi^0$&${1\over9}(D+3F^2)\bp$\\[2mm]
$\Sigma^+$&$-{2\over9}D^2(\bp+3\bm)$\\[2mm]
$\Sigma^-$&$-{2\over9}D^2(\bp-3\bm)$\\[2mm]
$\Lambda$&${2\over9}D^2\bp$\\[2mm]
$\Sigma^0\Lambda$&${2\over3\sqrt{3}}D^2\bp$\\[2mm]
\hline
\end{tabular}
\end{center}

\begin{center}
\begin{tabular}{|l|c|c|c|}\hline
&\multicolumn{3}{|c|}{$\tilde{\gamma}^{(X)}_B$}\\[1mm]\hline
&$\pi$&$K$&$\eta$\\[1mm]\hline
& & &\\[1mm]
$p$&${20\over27}\muC\c^2$&${5\over54}\muC\c^2$&$0$\\[2mm]
$n$&$-{5\over27}\muC\c^2$&$-{5\over54}\muC\c^2$&$0$\\[2mm]
$\Xi^-$&$-{5\over108}\muC\c^2$&$-{10\over27}\muC\c^2$&$-{5\over36}\muC\c^2$\\[2mm]
$\Xi^0$&$-{5\over54}\muC\c^2$&$-{5\over27}\muC\c^2$&$0$\\[2mm]
$\Sigma^+$&${5\over108}\muC\c^2$&${35\over54}\muC\c^2$&${5\over36}\muC\c^2$\\[2mm]
$\Sigma^-$&$-{5\over108}\muC\c^2$&$-{10\over27}\muC\c^2$&$-{5\over36}\muC\c^2$\\[2mm]
$\Lambda$&$0$&$-{5\over36}\muC\c^2$&$0$\\[2mm]
$\Sigma^0\Lambda$&${5\over18\sqrt{3}}\muC\c^2$&${5\over36\sqrt{3}}\muC\c^2$&$0$\\[2mm]
\hline
\end{tabular}
\end{center}

\begin{center}
\begin{tabular}{|l|c|c|c|}\hline
&\multicolumn{3}{|c|}{$\hat{\gamma}^{(X)}_B$}\\[1mm]\hline
&$\pi$&$K$&$\eta$\\[1mm]\hline
& & &\\[1mm]
$p$&${4\over9}(D+F)\c\muT$&${1\over9}(3D-F)\c\muT$&$0$\\[2mm]
$n$&$-{4\over9}(D+F)\c\muT$&$-{2\over9}F\c\muT$&$0$\\[2mm]
$\Xi^-$&${2\over9}(F-D)\c\muT$&${1\over9}(F-D)\c\muT$&$0$\\[2mm]
$\Xi^0$&${2\over9}(F-D)\c\muT$&${2\over9}(D+2F)\c\muT$&$-{1\over9}(D+3F)\c\muT$\\[2mm]
$\Sigma^+$&${1\over9}(D+3F)\c\muT$&${4\over9}D\c\muT$&${2\over9}D\c\muT$\\[2mm]
$\Sigma^-$&${1\over9}(F-D)\c\muT$&${2\over9}(F-D)\c\muT$&$0$\\[2mm]
$\Lambda$&$-{1\over3}D\c\muT$&${1\over9}(D-3F)\c\muT$&$0$\\[2mm]
$\Sigma^0\Lambda$&${1\over18\sqrt{3}}(D+6F)\c\muT$&${2\over9\sqrt{3}}(2D+3F)\c \muT$&${1\over6\sqrt{3}}D\c\muT$\\[2mm]
\hline
\end{tabular}
\end{center}

\begin{center}
\begin{tabular}{|l|c|c|c|}\hline
&\multicolumn{3}{|c|}{$\lambda^{(X)}_B$}\\[1mm]\hline
&$\pi$&$K$&$\eta$\\[1mm]\hline
& & &\\[1mm]
$N$&${9\over4}(D+F)^2$&${5\over2}D^2-3DF+{9\over2}F^2$&${1\over4}(D-3F)^2$\\[2mm]
$\Xi$&${9\over4}(D-F)^2$&${5\over2}D^2+3DF+{9\over2}F^2$&${1\over4}(D+3F)^2$\\[2mm]
$\Sigma$&$D^2+6F^2$&$3(D^2+F^2)$&$D^2$\\[2mm]
$\Lambda$&$3D^2$&$D^2+9F^2$&$D^2$\\[2mm]
$\Sigma^0\Lambda$&$2D^2+3F^2$&$2D^2+6F^2$&$D^2$\\[2mm]
\hline
\end{tabular}
\end{center}
\begin{center}
\begin{tabular}{|l|c|c|c|}\hline
&\multicolumn{3}{|c|}{$\tilde{\lambda}^{(X)}_B$}\\[1mm]\hline
&$\pi$&$K$&$\eta$\\[1mm]\hline
& & &\\[1mm]
$N$&$2\c^2$&${1\over2}\c^2$&$0$\\[2mm]
$\Xi$&${1\over }\c^2$&${3\over2} \c^2$&${1\over2}\c^2$\\[2mm]
$\Sigma$&${1\over3} \c^2$&${5\over3} \c^2$&${1\over2} \c^2$\\[2mm]
$\Lambda$&${3\over2} \c^2$&$\c^2$&$0$\\[2mm]
$\Sigma^0\Lambda$&${11\over12}\c^2$&${4\over3}\c^2$&${1\over4} \c^2$\\[2mm]
\hline
\end{tabular}
\end{center}

\end{document}